# Fabrication and characterization of a Ni-Mn-Ga uniaxially textured freestanding film deposited by DC magnetron sputtering

**J. Tillier, D. Bourgault, B. Barbara, S. Pairis, L. Porcar, P. Chometon, D. Dufeu**

Institut Néel/Consortium de Recherche pour l'Emergence de Technologies Avancées, Centre National de la Recherche Scientifique, 38000 Grenoble, France

**N. Caillault, L. Carbone**

Schneider Electric Industries, 38000 Grenoble, France

**Corresponding author**

Name: Jérémy Tillier

Address: CRETA/CNRS, 25 avenue des martyrs, BP166 38042 Grenoble Cedex 09

Phone number: +33476889044

Fax number: +33476881280

E-mail address: jeremy.tillier@grenoble.cnrs.fr

**Abstract**

Homogeneous freestanding films have been obtained by the direct current (DC) magnetron sputtering technique using a sacrificial layer. After annealing, the films are crystallized with a strong out-of-plane texture along the (022) direction. The stoichiometry of the annealed films is close to the target composition and leads to a martensitic transformation around 255K. The annealed films demonstrate ferromagnetic behavior with a Curie temperature of about 362K. The magnetization process has been studied on the both states and during the martensitic transition. The saturation magnetizations have been determined by fitting the experimental data with a saturation approach law in the range 1-5T. Results show the saturation magnetization of the martensite is around 10% higher than that of the austenite. A model based on intrinsic magnetic properties of each state allowing the description of the magnetization M=f (H, T) of such polycrystalline films during the martensitic transformation is presented. The mass fraction of martensite inside the austenite phase can be determined using this model. The shape memory effect is analyzed both by scanning electron microscopy and by optical microscopy with in-situ measurement of the resistance temperature dependence.

**1. Introduction**

The ternary intermetallic compound $Ni_2MnGa$ has gained considerable attention, as it exhibits an interesting combination of thermoelastic and magnetic properties. In addition to the conventional shape memory effect, this system shows a magnetic field induced rearrangement of martensitic variants commonly called as magnetic shape memory (MSM) effect [1-4]. Recent investigations in $Ni_2MnGa$ bulk single crystals revealed considerable shape changes up to 10% and an interesting response time below 1 ms [6-11]. So Ni-Mn-Ga alloys have a great potential application as microactuators and microsensors for micro-electro-mechanical systems.

However, the bulk alloy is too brittle to be machined in a required shape and Ni-Mn-Ga alloys have to be prepared directly in film form [12]. So far, Ni-Mn-Ga thin films have been grown by means of various techniques such as molecular beam epitaxy [13], sputtering [14-22] and pulsed laser deposition [23-24]. Freestanding Ni-Mn-Ga thin films with conventional shape memory effect have already been implemented. Recent studies on freestanding epitaxially grown Ni-Mn-Ga thin film actuators indicate magnetic shape memory although no quantitative results have been obtained [13]. For freestanding polycrystalline thick films, the intrinsic properties still remain largely unexplored.

In this paper, the fabrication process of freestanding uniaxially textured Ni-Mn-Ga films is described in detail. The martensitic transformation and the magnetic properties of a freestanding polycrystalline $Ni_{52.5}Mn_{24}Ga_{23.5}$ thick film are studied with respect to its texture. The shape memory effect is observed during the austenitic transformation by scanning electron microscopy. The full austenite to martensite and martensite to austenite shape memory cycle is analyzed by optical microscopy with in-situ measurement of the resistance vs. temperature curve.

2. Experimental

Ni-Mn-Ga thick films have been deposited on photoresist (Shipley S1818) sacrificial layers, which were spin-coated on polycrystalline alumina substrates. The Ni-Mn-Ga films with a thickness of about 5 μm were deposited using the direct current (DC) magnetron sputtering technique using a two-inch target of a ternary Ni-Mn-Ga alloy. The sputtering conditions were as follows: the base vacuum in the sputtering chamber was below $1.10^{-3}$ Pa. The sputtering power was 25W and the distance between the substrate and the target was 5.5 cm. The argon gas with a high purity of 99.999 vol% was filled into the chamber to keep a working pressure of 0.5 Pa. The sputtering duration was 300 min.

After deposition, the Ni-Mn-Ga/S1818/Al$_2$O$_3$ composite was placed into an acetone bath to strip the resist sacrificial layer. Then, the freestanding film fastened by alumina ceramic pieces was annealed at 1093K for 6h under a vacuum below 1.10$^{-3}$ Pa followed by furnace-cooling to achieve homogenization and ordering.

The crystallographic structure of the film was analyzed using an X-ray diffraction (XRD) instrument (Siemens D5000) with Cu-K$_\alpha$ radiation. The annealed sample for XRD analysis was a freestanding film pasted onto an amorphous glass piece. The in-plane magnetization loops were studied using a vibrating sample magnetometer (VSM Oxford). The measurements were made at 220 and 295K, in the martensitic and the austenitic states, respectively. The magnetization was then measured from 5T to 0T in the range 226-280K for selected values of the temperature. The magnetic transition behavior was determined by the temperature dependence of the magnetization at 0.2T using an extraction magnetometer. Measurements were done on heating from 290K up to 400K. The microstructure of the annealed films was examined by scanning electron microscopy (SEM JEOL 840A). In order to observe the shape memory effect, the SEM sample holder was cooled down by liquid nitrogen circulation. The optical observations were realized using a Zeiss microscope equipped with a Hamamatsu ORCA-ER digital camera. The sample was cooled down with a hand-made liquid helium circulation cryostat with in-situ measurement of the resistance by the four probe method [25].

3. Results and discussion

*3.1. Structural properties*

The annealing parameters allow a complete crystallization of the film. After annealing at 1093K for 6h, the XRD pattern of the annealed film shown in fig. 1 can be indexed by the Heusler-type Fm-3m cubic austenitic structure with a lattice parameter of 5,797 Å. In this cubic L2$_1$ ordered phase only the (022) and (044) peaks were found. The film demonstrates a strong (022) out-of-plane texture. The XRD pattern did not show additional peaks, hence no precipitation or formation of others phases occurs. The SEM observation, shown in fig. 2, displays an in-plane polygonal grain structure with an average size of 7 μm. The composition is homogenous across the film and was determined to be $Ni_{52.5}Mn_{24}Ga_{23.5}$ (at %) by EDX. The contrast seen on the backscattered electron SEM image might be due to different in-plane crystallographic orientation in each grain. Four-circle XRD texture measurements and electron backscattered diffraction (EBSD) analysis will be done to confirm this observation.

### *3.2. Magnetic properties*

Fig. 3 shows the temperature dependence of the magnetization in the range 226-400K under an in-plane applied magnetic field of 0.2T. This curve reveals the Curie temperature $T_C$ of the film is approximately 362K. The magnetization jump around 255K corresponds to the martensitic transition. The characteristics temperatures of this first-order magnetic phase transition were determined by a conventional tangential method. The austenite to martensite transformation begin at $M_S$=261K and finishes at $M_F$=250K. The film stoichiometry leads to a valence electron density e/a of 7.635. The measured martensite start temperature is consistent with that reported by Chernenko et al. for bulk samples exhibiting the same e/a [15, 26].

Fig. 4 displays two magnetization loops of the film measured on the both sides of the martensitic transformation. The magnetic field was applied parallel to the film surface. In this configuration, no correction is needed because the demagnetization field can be neglected. The high temperature cubic austenite phase demonstrates no coercivity. The magnetization increases by magnetic domain wall motion and saturates in a very low applied field of around 0.08T. The transition to martensite phase causes distortion of the crystal structure lattice, leading to increasing of the magnetocrystalline anisotropy and hardening of the magnetic saturation process. The film exhibit a strong (022) out-of-plane texture. When the film is cooled down through the transition, two variants with their easy axis forming a 45° angle with the film surface can nucleate. The possibly third variant have his easy axis in the plane of the film but not necessary in the field direction because of the absence of in-plane texture. The magnetization loop is the sum of the magnetization loops of each crystallite where three types of variants can nucleate with the orientations described above. The magnetic domain wall nucleation and propagation inside the grains takes place at the beginning. Then, the magnetization rotation process inside the magnetic domains occurs for all the grains with their easy axes not parallel to the applied magnetic field. The saturation field of the martensite, much more intense than that of the austenite phase, is around 1.3T.

Fig. 5 shows the temperature dependence of the magnetization measured for the martensitic transformation on cooling, for selected values of the in-plane magnetic field. The magnetization M displays a significantly abrupt change ΔM = M(austenite) − M(martensite) at the phase transition. In the low-field range, the magnetization displays a sharp decline which corresponds to the increase of magnetocristalline anisotropy in the martensitic state. At strong fields, the temperature dependence of the magnetization shows a slight slope increase during the martensitic transformation. It indicates a higher saturation magnetization in the martensite state. It is due to the magnetostructural coupling between magnetic moments and variants in the twinned martensite structure [27]. In other Heusler system like Ni-Mn-Sn or Ni-Mn-In, the coupling between structure and magnetism lead to a lower saturation magnetization of the martensite [28]. The magnetoelastic interaction is also responsible of the large strain caused by the field induced rearrangement of martensitic variants in single Ni-Mn-Ga single crystals [29].

At intense fields, the law of the saturation approach is given by:

$$M_H = M_S \left[1 - \frac{a}{H}\right] + \chi H \quad (1)$$

The superposed susceptibility $\chi$ is dominated by the Pauli term of conduction electrons. The term $\frac{a}{H}$ is attributed by Néel to the effects of defects in crystallites and especially at grain boundaries. $M_H$ is the magnetization at an intense field $H$ and $M_S$ is the saturation magnetization. In order to estimate the saturation magnetization difference between the two states, each magnetization curve was fit using equation (1) in the range 1-5T. Agreement between experimental points and the fit is excellent. The $M_H - \chi H = f(\frac{1}{H})$ representations are perfectly linear (not shown). Fig. 6 shows the temperature dependence of the fit parameter $a$ and the superposed susceptibility $\chi$. Each parameter demonstrates a fast evolution when the martensite nucleates in the parent austenitic phase. The parameter $a$ shows a strong increase due to the much harder magnetization process of the martensite while the superposed susceptibility $\chi$ decreases drastically indicating a smaller conduction electron density in the martensitic state. It is consistent with the temperature dependence of the electrical resistance reported in the literature, the transition to martensite resulting in a fast increase of the electrical resistance [14, 16-18]. In the electron gas model, the resistivity $\rho$ is given by:

$$\rho = \frac{m}{Ne^2 \tau} \quad (2)$$

The resistivity depends on the electron density $N$ and the relaxation time $\tau$ between two collisions. The constants $m$ and $e$ are the mass and the elementary charge of the electron, respectively. The resistivity being inversely proportional to the electron density, she increases when the electron density is decrease. Thus, the evolution observed at the martensitic transformation for the superposed susceptibility $\chi$ is consistent with the fast resistance increase displayed on Fig. 10 and reported in the literature [14, 16-18]. Fig. 7 represents the field dependence of the magnetization analyzed from 2T to 0T for selected values of the temperature. Two curves have been measured in the austenitic state by controlling the temperature at 275 and 280K. The second pair of curve corresponds to the martensitic state, at 226 and 241K. The last curve has been measured at 255K, during the transition. It displays the magnetic saturation process of a mix of austenite and martensite which can be described by a mass fraction of martensite $\alpha^M$. Assuming that the mass fraction $\alpha^M$ doesn't show magnetic field dependence (see Fig. 5) and including this mass fraction of martensite $\alpha^M$, the magnetization during the transition can be described as follows:

$$M_{(H,T)} = \alpha^M_{(T)} . M^M_{(H,T)} + (1 - \alpha^M_{(T)}) . M^A_{(H,T)} \qquad (3)$$

In the above equation, the magnetization $M_{(H,T)}$ is a function of the magnetic field $H$ and the temperature $T$. The mass fraction of martensite $\alpha^M_{(T)}$ depends only on the temperature $T$. $M^M_{(H,T)}$ and $M^A_{(H,T)}$, which are temperature and field dependent, are the magnetization of the martensitic phase and the austenitic one, respectively. The martensitic transformation occurs far from the ferromagnetic to paramagnetic transition and takes place in a small temperature range. The effect of the temperature on the magnetization for a fixed field can thus be neglected during the transformation. The temperature at which the mass fraction of martensite $\alpha^M$ = 0% and $\alpha^M$ = 100% have been determined using the fit parameters evolutions (see Fig. 6). With respect to the above mentioned approximation, equation (3) becomes:

$$M_{(H,T)} = \alpha^M_{(T)} M^M_{(H,245\,K)} + (1 - \alpha^M_{(T)}) M^A_{(H,265\,K)} \qquad (4)$$

To determine the temperature dependence of the martensite mass fraction $\alpha^M$ during the transition, the magnetization curves measured in the temperature range of the transition have been fitted with equation (4). Fits of experimental data for selected values of temperature are depicted in fig. 8. It demonstrates the excellent agreement between experience and model. To improve the fit quality, the effect of the temperature on the magnetization for a fixed field can be included using a classical Langevin law. Evolutions of the saturation magnetization $M_S$ and the martensite mass fraction $\alpha^M$ with the temperature are represented in fig. 9. For the both states, the saturation magnetizations follow a Langevin law, resulting in a magnetization increase for decreasing temperatures. A slope increase of the saturation magnetization can be seen during the martensitic transition. In fact, the martensite mass fraction increases drastically during this first order magnetic phase transition. The martensite having a saturation magnetization slightly higher than the austenite, it results in a slope change of the saturation magnetization of the phase mix during the martensitic transformation. By extrapolation of the saturation magnetization in the parent austenite phase, the saturation magnetization of the martensite is found to be around 10% larger than the austenite one (see Fig. 9).

Whatever the temperature, no evidence of field-induced rearrangement of twin variants is observed (not shown). It is imputable to the polycrystalline nature and the absence of in-plane texture of the films. Our group is now focusing on the film texturation in order to obtain biaxially oriented film.

### 3.3. *Shape memory effect*

The temperature-dependent local morphology of a 0.14 mm$^2$ area studied with in-situ four-probes measurement of the resistance is shown in Fig. 10. The observation area was located between the two inner voltage measurement probes. A video clip of the complete transformation is available online. Within the austenitic state (Fig. 10 a and f) the surface is flat and shiny. When the film is cooled down through the transition, some areas with a different roughness state appear. The observed temperature-dependent appearance of rough areas is ascribable to martensitic grains originating from different nucleation sites which grow at the expense of austenite. By heating the film from the martensitic state, austenitic grains grow in the martensite and the film finds again his flat surface with a shinny appearance. The roughness appearance in the martensitic state can be explained by the mesoscopic shape memory effect. The austenitic grain transforming in martensite demonstrates mesoscopic shape changes due to the cubic austenite to lower symmetry martensite transition. The film being out-of-plane texture with no in-plane texture, each grain deforms in random directions leading to crinkle the freestanding film in some grains or at grain boundaries. It can be seen on Fig. 11 which represents the temperature-dependent local SEM morphology of a 0.017 $mm^2$ area during the austenitic transformation. The much better field depth of the SEM compared to optical microscopy allows focalizing the entire film surface despite of the strong surface roughness in the martensitic state. Fig. 11 demonstrates the appearance of relief, the film crinkling at some grain boundaries and especially in the grains. The temperature dependence of the resistance displayed in Fig. 10 shows that the transformation process of the film is not fully finished when the observed area is completely transformed. In fact, the observed area is fully martensitic at 257K, well above the end of the transition (see the resistance vs. temperature curve on Fig. 10). For the reverse transformation, the observed area is fully martensitic well above the beginning of the transformation, the martensitic to austenite transition occurring first in another area of the film. It agrees with a first-order phase transition, different nucleus growing at the expense of the other phase. The freestanding Ni-Mn-Ga film exhibits a fully reversible shape memory surface. It shows two states with an important roughness difference. A strong reflective index jump is expected. It is due to mesoscopic shape memory effects during the austenite to martensite and the reverse transformations. Thus, the freestanding Ni-Mn-Ga film might be a good candidate for micro-opto-electro-mechanical systems.

4. **Summary and conclusions**

Freestanding Ni-Mn-Ga thick films have been successfully prepared by the DC magnetron sputtering technique using a sacrificial layer. After heat-treatment, the $Ni_{52.5}Mn_{24}Ga_{23.5}$ (at %) freestanding films present a strong (022) out-of-plane texture, display martensitic transformation and first order ferromagnetically with a Curie temperature of around 362K. By using a saturation approach law, the saturation magnetization of the martensite has been found to be 10% higher than that of the austenite. A model based on intrinsic magnetic properties allowing the description of the magnetization M=f (H, T) of such polycrystalline films have been presented. The mass fraction of martensite inside the austenite phase can be determined using this model. The shape memory effect of such freestanding uniaxially textured Ni-Mn-Ga films has been analyzed. The film exhibits a fully reversible shape memory surface with two roughness states. In the austenitic state, the surface is flat and shinny. When the film is cooled down through the martensitic transition, a strong relief appears, due to mesoscopic shape memory effects in each grain.

**Acknowledgements**

The NanoFab and magnetometry poles of the nanoscience department (Institut Néel, CNRS, Grenoble, France) are gratefully acknowledged for their contribution in this work.

**References**

[1] O. Söderberg, Y. Ge, A. Sozinov, S.-P. Hannula, V.K. Lindroos, Smart Mater. Struct. 14 (2005) 223

[2] O. Söderberg, I. Aaltio, Y. Ge, O. Heczko, S.-P. Hannula, Materials Science and Engineering A 481-482 (2008) 80

[3] B. Winzek, S. Schmitz, H. Rumpf, T. Sterzl, R. Hassdorf, S. Thienhaus, J. Feydt, M. Moske, E. Quandt, Materials Science and Engineering A 378 (2004) 40

[4] V.A. Chernenko, S. Besseghini, Sensors and Actuators A 142 (2008) 542

[5] K. Ullakko, J.K. Huang, C. Kantner, R.C. O'Handley, V.V. Kokorin, Appl. Phys. Lett. 69 (1996) 1966

[6] A.A. Likhachev, K. Ullakko, Physics Letters A 275 (2000) 142

[7] S.J. Murray, M. Marioni, S.M. Allen, R.C. O'Handley, T.A. Lograsso, Appl. Phys. Lett. 77 (2000) 886

[8] O. Heczko, A. Sozinov, K. Ullakko, IEEE Transactions on Magnetics 32 (2000) 3266

[9] R.D. James, R. Tickle, M. Wuttig, Materials Science and Engineering A 273-275 (1999) 320

[10] A. Sozinov, A.A. Likhachev, N. Lanska, K. Ullakko, Appl. Phys. Lett. 80 (2002) 1746

[11] P. Müllner, V.A. Chernenko, G. Korstorz, J. Appl. Phys. 95 (2004) 1531

[12] H.B. Xu, Y. Li, C.B. Jiang, Materials Science and Engineering A 438-440 (2006) 1065

[13] J.W. Dong, J.Q. Xie, J. Lu, C. Adelmann, C.J. Palmstrom, J. Cui, Q. Pan, T.W. Shield, R.D. James, S. McKernan, J. Appl. Phys. 95 (2004) 2593

[14] V.A. Chernenko, S. Besseghini, M. Hagler, P. Müllner, M. Ohtsuka, F. Stortiero, Materials Science and Engineering A 481-482 (2008) 271

[15] M. Ohtsuka, M. Matsumoto, K. Itagaki, J. Intelligent Material Systems and Structures 17 (2006) 1069

[16] M. Kohl, D. Brugger, M. Ohtsuka, T. Takagi, Sensors and Actuators A 114 (2004) 445

[17] S. Besseghini, A. Gambardella, V.A. Chernenko, M. Hagler, C. Pohl, P. Müllner, M. Ohtsuka, S. Doyle, Eur. Phys. J. Special Topics 158 (2008) 179

[18] V. Chernenko, M. Kohl, S. Doyle, P. Müllner, M. Ohtsuka, Scripta Materiala 54 (2006) 1287

[19] G. Jakob, H.J. Elmers, Journal of Magnetism and Magnetic Materials 310 (2007) 2779

[20] M. Thomas, O. Heczko, J. Buschbeck, L. Schultz, S. Fähler, Appl. Phys. Lett. 95 (2008) 192515

[21] F. Khelfaoui, M. Kohl, J. Buschbeck, O. Heczko, S. Fähler, L. Schultz, Eur. Phys. J. Special Topics 158 (2008) 167

[22] O. Heczko, M. Thomas, J. Buschbeck, L. Schultz, S. Fähler, Appl. Phys. Lett. 92 (2008) 072502

[23] P.G. Tello, F.J. Castano, R.C. O'Handley, S.M. Allen, M. Esteve, A. Labarta, X. Batlle, J. Appl. Phys. 91 (2002) 8234

[24] A. Hakola, O. Heczko, A. Jaakkola, T. Kajava, K. Ullakko, Applied Surface Science 238 (2004) 155

[25] A. Villaume, A. Antonevici, D. Bourgault, J.P. Leggeri, L. Porcar, C. Villard, Review of Scientific Instruments 79 (2008) 023904

[26] V.A. Chernenko, Scripta Materiala 40 (1999) 523

[27] J. Marcos, L. Manosa, A. Planes, F. Casanova, X. Batlle, A. Labarta, Physical Review B 68 (2003) 094401

[28] X. Moya, L. Manosa, A. Planes, T. Krenke, M. Acet, E. Wassermann, Materials Science and Engineering A 438-440 (2006) 911-915

[29] V.A. Chernenko, V.A. L'vov, P. Müllner, G. Kostorz, T. Takagi, Physical Review B 69 (2004) 134410

*Fig.1*

Θ-2Θ X-ray diffraction spectrum of the annealed film

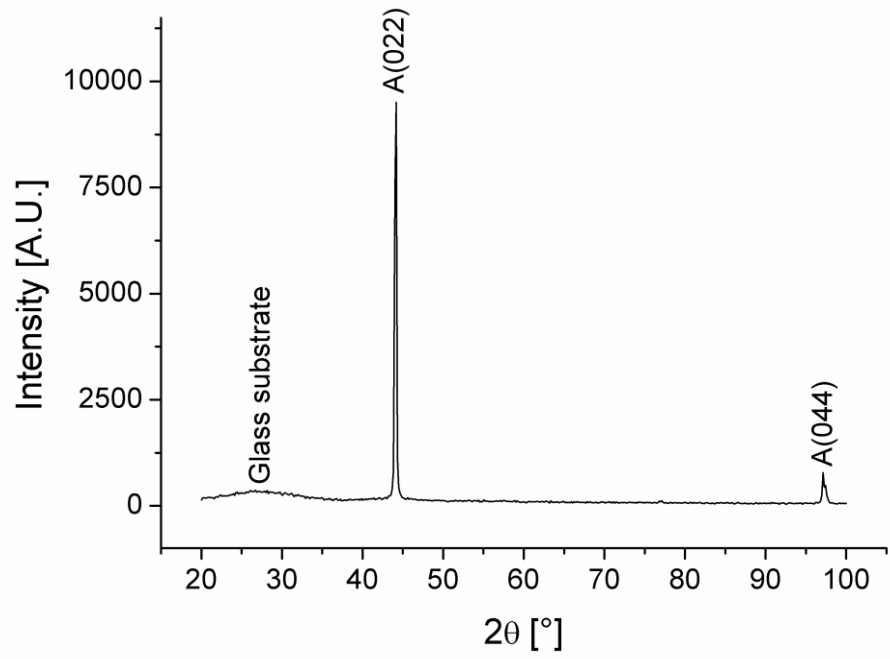

*Fig.2*

Typical backscattered electron SEM image of the surface morphology for the annealed film.

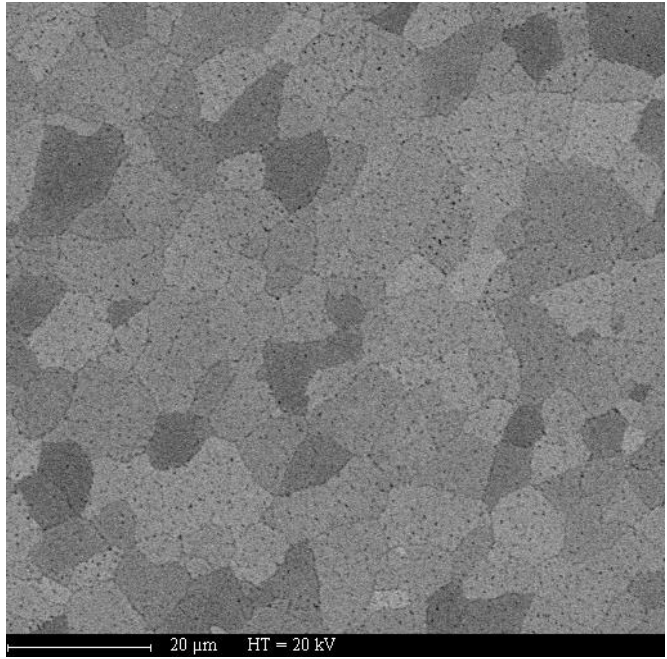

*Fig.3*

Temperature dependence of the magnetization for the annealed film under an in-plane applied magnetic field of 0.2T.

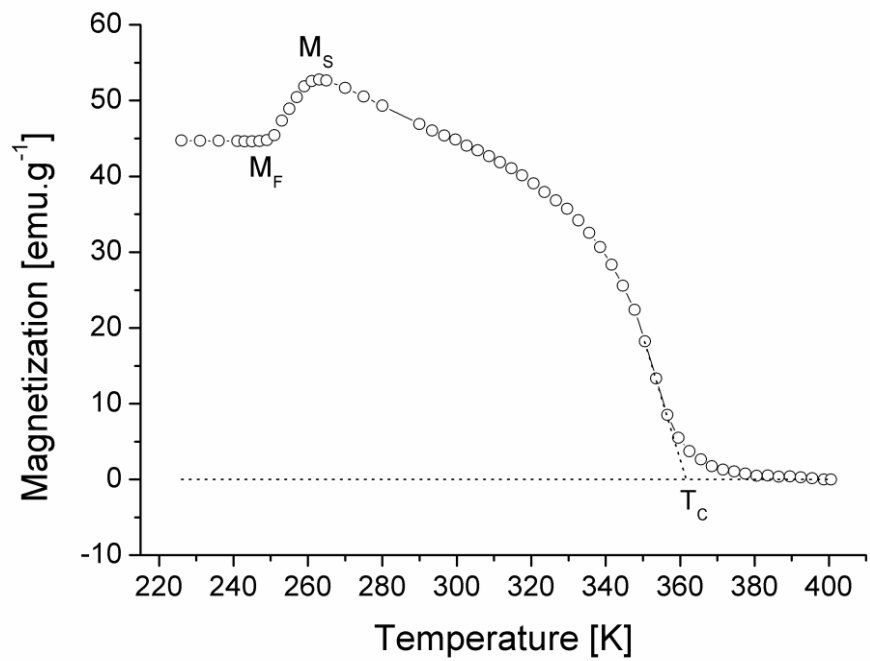

*Fig.4*

Magnetization loops of the annealed film for selected temperature in the martensitic and the austenitic states.

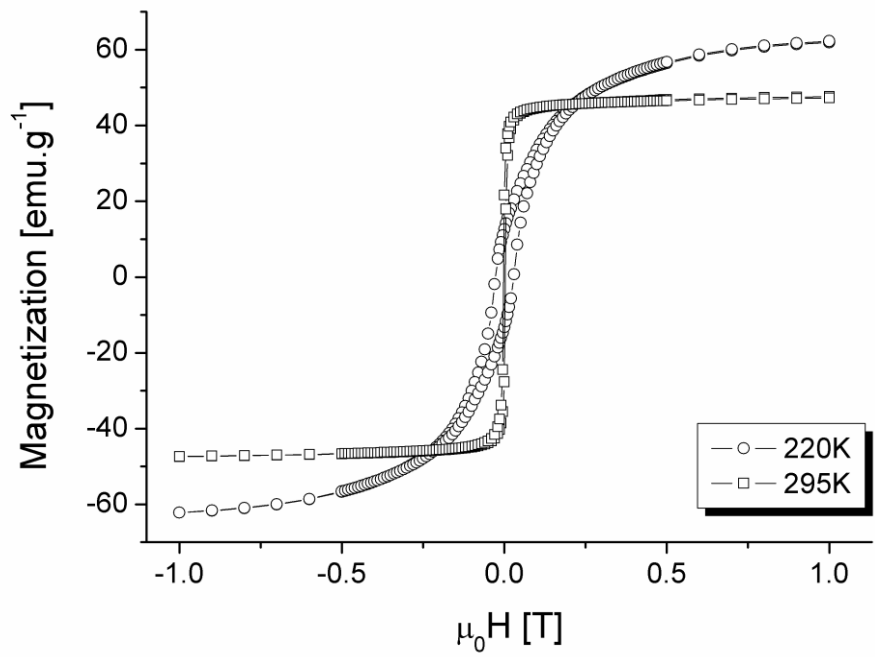

*Fig.5*

Temperature dependence of the magnetization for selected values of the in-plane applied magnetic field. The measurements were performed on cooling.

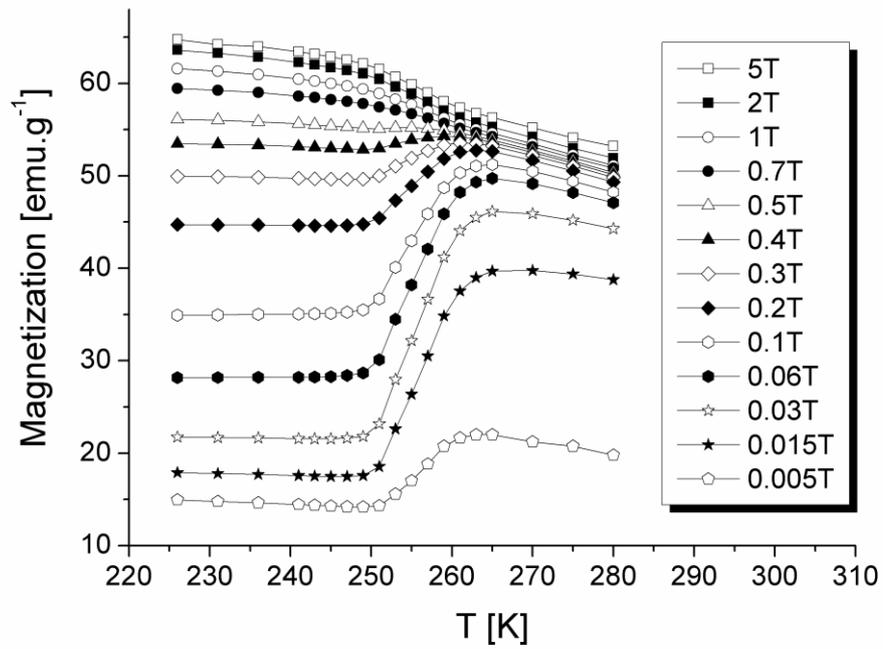

*Fig.6*

Temperature dependences of the fit parameters $a$ (open down triangle) and $\chi$ (open star).

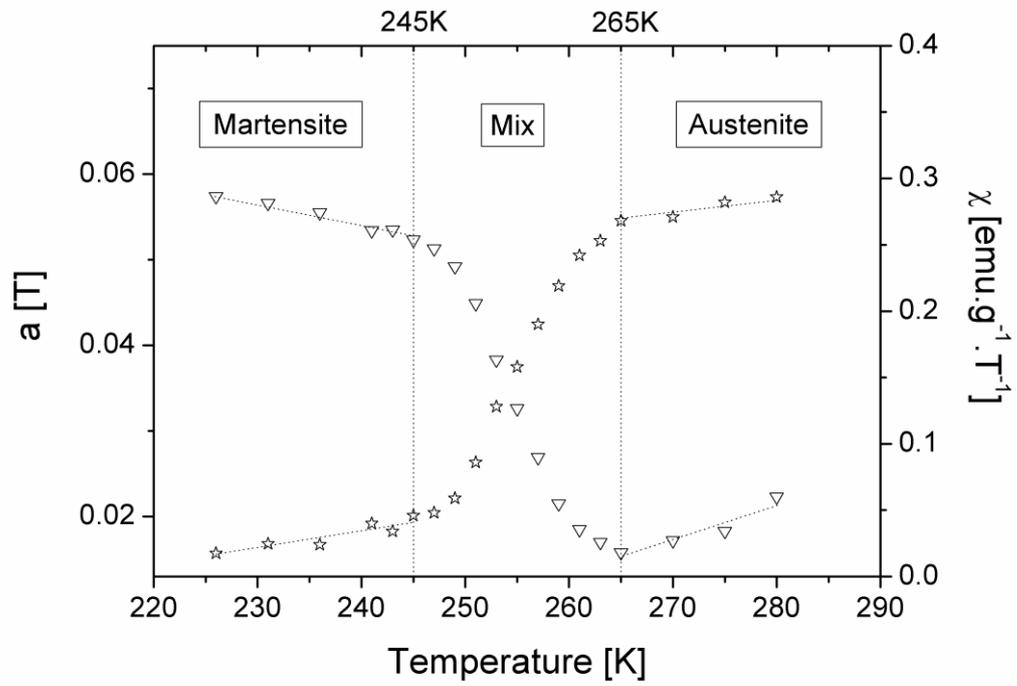

*Fig.7*

The magnetization process measured from 2 to 0T for selected values of the temperature. The half-up diamond corresponds to a temperature at which the two phases cohabit.

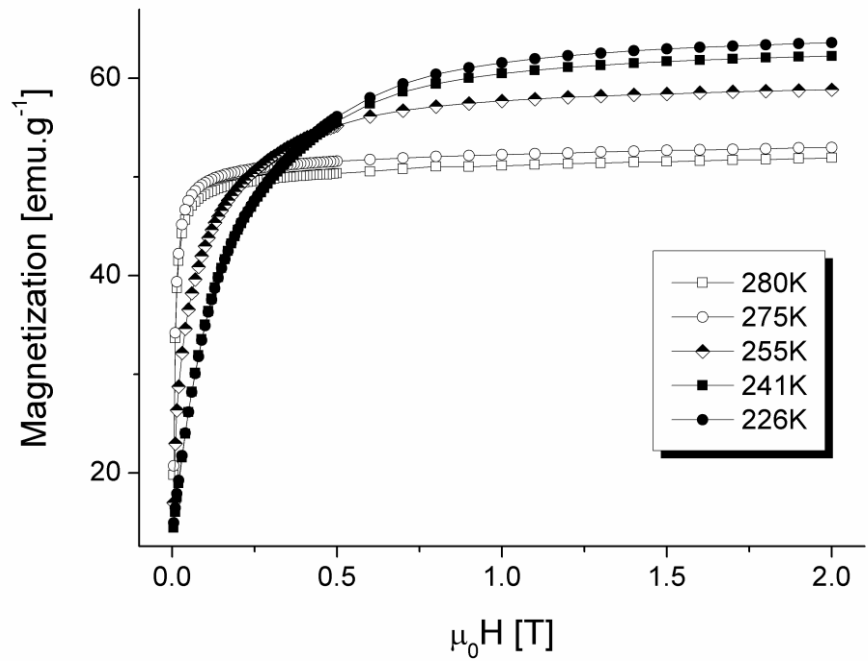

*Fig.8*

Experimental M=f (H) points fitted with the model proposed at equation (4) for selected values of the temperature. The corresponding mass fractions of martensite are indicated in the legend.

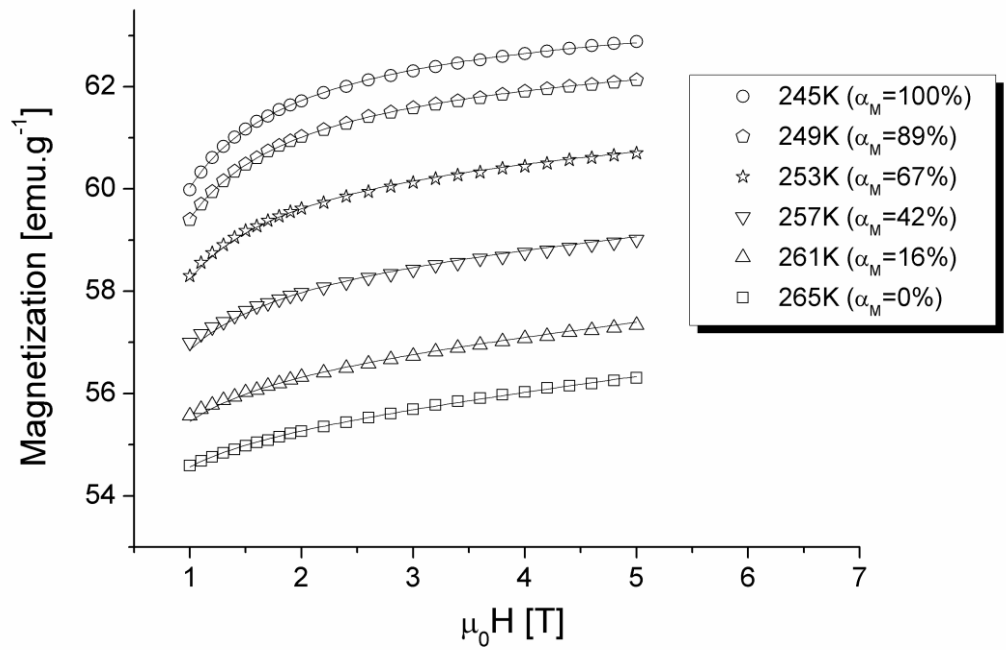

*Fig.9*

Temperature dependences of the martensite mass fraction $\alpha_M$ and the magnetization at saturation $M_S$. By extrapolation, the saturation magnetization of the martensite is found to be around 10% higher than the austenite one.

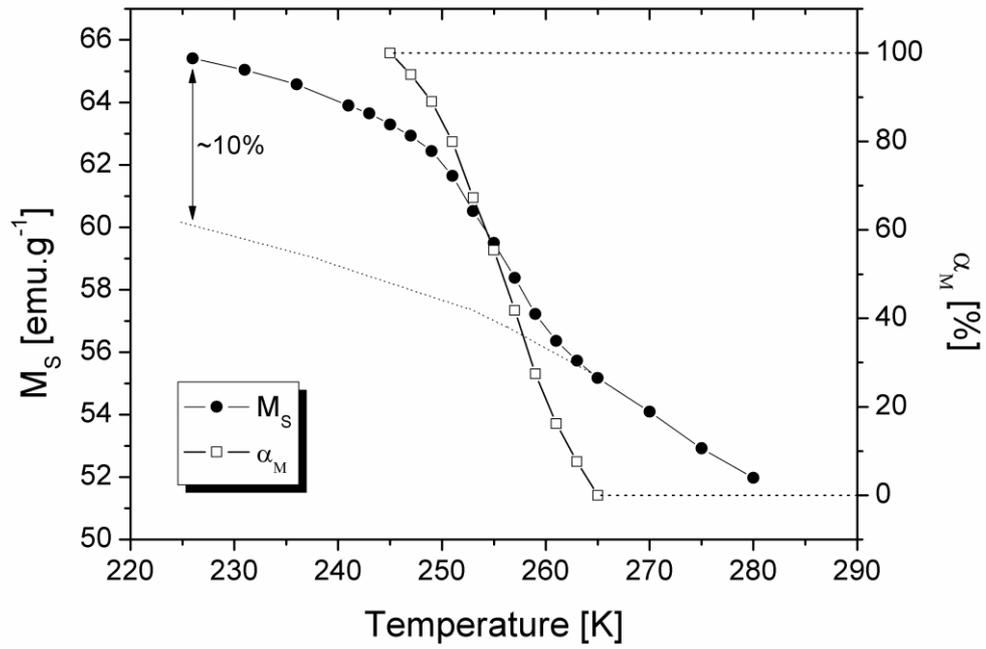

*Fig.10*

Optical micrographs of the surface morphology with in-situ measurement of the resistance temperature dependence. The arrows on the resistance vs. temperature curve indicate the temperatures at which the presented images were done.

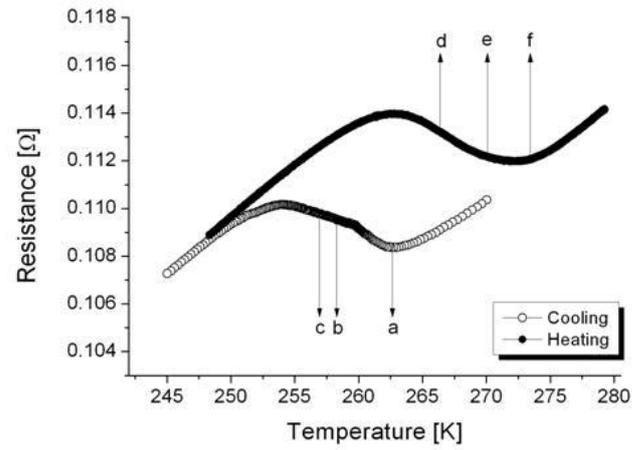

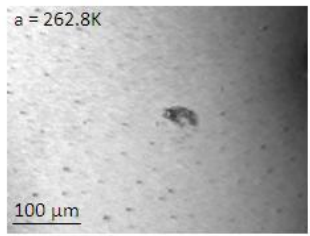 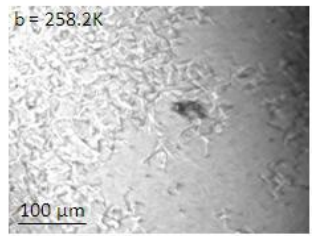 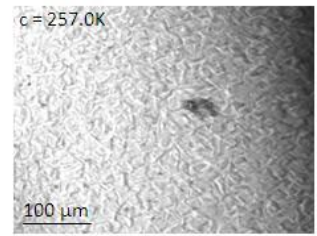

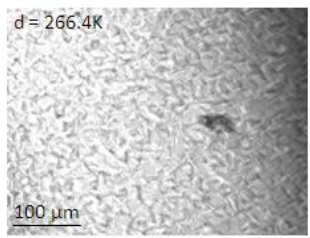 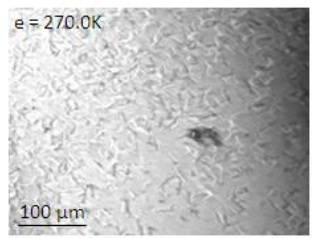 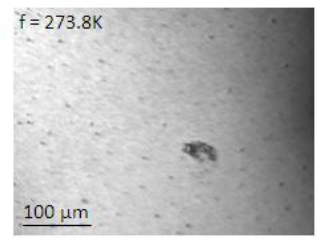

*Fig. 11*

Typical secondary electron SEM images of the surface morphology for increasing temperatures from the martensitic state (a) to the austenitic one (c).

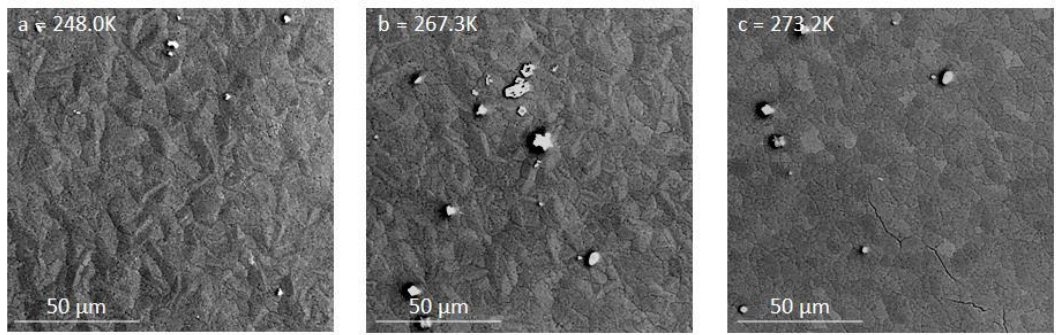